\documentclass[twocolumn, final]{svjourCustomized3} %, global, ,referee
\usepackage{latexsym}
\usepackage{graphics}
\usepackage{cite}
\usepackage{color}
\usepackage{hyperref}
\usepackage{layouts}

\usepackage[normalem]{ulem} %# to use \sout
\begin{document}
\sloppy 
%\sloppy  as suggested in http://tex.stackexchange.com/questions/32062/automatic-line-breaking-for-two-column-text
%
\journalname{\ }
\date{\ }
\title{Towards a quantum interface between telecommunication and UV wavelengths: design and classical performance}
\titlerunning{Towards a quantum interface between telecommunication and UV wavelengths}
\author{Helge R\"utz,  Kai-Hong Luo, Hubertus Suche, Christine Silberhorn }
\institute{Integrierte Quantenoptik, Universit\"at Paderborn, Warburger Stra{\ss}e 100, 33098 Paderborn, Germany\\\email{helge.ruetz@uni-paderborn.de}}
%\href{helge.ruetz@uni-paderborn.de}{\nolinkurl{helge.ruetz@uni-paderborn.de}}
%

\maketitle

%
%\address{$^1$ Integrierte Quantenoptik, Universit\"at Paderborn, Warburger Stra{\ss}e 100, 33098 Paderborn, Germany}
%\email{$^*$helge.ruetz@uni-paderborn.de} %% email address is required
%
%
%%%%%%%%%%%%%%%%%%% abstract and codes %%%%%%%%%%%%%%%%
%
%
\begin{abstract}
We propose and characterize a quantum interface between telecommunication wavelengths (1311~nm) and an Yb${}^{+}$-dipole transition (369.5~nm) 
 based on a second order sum frequency process in a \mbox{PPKTP} waveguide. An external (internal) conversion efficiency above~5\%~(10\%) is shown using classical bright light.  
\end{abstract}
\PACS{42.79.Nv \and 42.82.-m 	\and 42.65.Wi}
%
%42.79.Nv 	Optical frequency converters \and 
%42.82.-m 	Integrated optics
%42.65.Wi 	Nonlinear waveguides
%
%%%%%%%%%%%%%% Finding out the Page style %%%%%%%%%%
%\thefontsize\footnotesize
%\thefontsize\small
%\thefontsize\Large
%\thefontsize\LARGE
%\showthe\font
%
%\drawmarginparstrue
%\currentpage
%\pagedesign
%
%%%%%%%%%%%%%% Start the paper %%%%%%%%%%
\section{Introduction}
A quantum interface (QI) is a basic building block for future quantum networks~\cite{Kimble2008}.
It allows to combine different physical systems without changing their shared quantum properties and may e.g. be used to change the color of a photonic quantum state while preserving its other quantum properties in order to combine atomic quantum memories and fiber communication channels. 

In such a network, the wavelengths of optimal transmission in optical fibers, situated in the infrared spectral range at telecommunication wavelengths, 
may differ significantly from the wavelengths at which quantum information 
can be generated, stored or processed locally. The latter frequently lie in the ultraviolet (UV) and blue spectral 
range~\cite{Olmschenk2010}.
Thus, to bridge this frequency gap between telecommunication and UV wavelengths, a QI with high efficiency, compact design and network compatibility is highly desirable for practical applications.

The most efficient process allowing to build such a QI is known as quantum frequency conversion (QFC)~\cite{Kumar1990}. 
Important initial work~\cite{Kumar1992, Vandevender2004, Albota2004} has lead to the realization of
quantum state preserving converters~\cite{Raymer2012}, showing the conversion of single photons in up-~\cite{Rakher2010, Ates2012} and down-conversion~\cite{Zaske2012, Ikuta2011} direction, as well as presenting the conversion of entanglement~\cite{Tanzilli2005, Ikuta2011} and squeezing~\cite{Vollmer2014, Kong2014}.
All of those experiments performed conversion between the green or red and the infrared spectral range. 
However, photonic state manipulation strongly benefits from de\-co\-her\-ence-free, high-energy transitions. Thus, trapped atoms and ions, whose dipole transitions usually lie in the ultraviolet (UV) and blue spectral range, are considered to be promising candidates for stationary qubit systems, allowing for long lived memories and high quality quantum logic gates.
The implementation of quantum frequency conversion between telecommunication and UV wavelengths is therefore 
an important step towards connecting dissimilar systems and paves the way for combining the possibilities of long distance quantum communication with the high fidelities achievable in atomic qubit systems. 

Still, the conversion of photonic qubits between telecommunication wavelengths and the UV is quite intricate, due to the special wavelength range and large frequency gap. Even in the classical regime, the generation of UV light by means of parametric three wave mixing processes is challenging. 
It has so far been realized through second harmonic generation (SHG)~\cite{Wang1998, Qing2001}, and sum frequency generation (SFG)~\cite{Oh1995, Corner2002, Berkeland1997, Umemura2003, Kumagai2003, Franzke1998}. 
While the SHG process is intrinsically not compatible with QFC, the realization of QFC via SFG imposes more stringent requirements on the process than classical wavelength conversion.
More specifically, a QFC interface based on a three wave mixing process in a nonlinear medium requires the high nonlinear coupling strength to be solemnly provided by the medium's nonlinearity together with a strong non-depleted pump. The pump wave thereby has to not only account for the energy difference between the converter's input and output wavelength, but also provide significant coupling strength to convert the faint input state.
The above SFG experiments fail to meet that requirement, by either falling short of the required coupling or by working with a depleted pump wave.
Hence, the underlying frequency conversion concepts cannot be readily applied to achieve QFC between telecommunication wavelengths and the UV spectral range.

Integrated optics provides a promising platform to implement a QI via QFC. 
Significant conversion efficiency can be achieved by using a waveguided converter~\cite{Roussev2004}. Although the required coupling strength can in principle be achieved by placing the nonlinear medium into a cavity~\cite{Vollmer2014, Kong2014}, the integrated optical approach allows for a robust and miniaturized system architecture that can be operated in single-pass configuration due to the high confinement obtained in the waveguide.

But in the case of an interface between the UV and telecommunication wavelengths the energy difference is extremely large, spanning more than $2.4\,\mathrm{eV}$ ($580\,\mathrm{THz}$). Bridging this huge gap either requires a multistage process, as proposed in~\cite{Clark2011}, or a strong pump in a waveguide with small poling period.
Many of the aforementioned quantum converters between visible and infrared make use of quasi-phasematching (QPM) in periodically poled lithium niobate (PPLN) waveguides.
However, those lithium niobate waveguides are highly susceptible to photorefractive damage which limits their use in conjunction with strong continuous wave (CW) green pump light.
Besides, to compensate the large wavelength difference and obtain efficient conversion, the QPM period (using the $d_{33}$ tensor element) in lithium niobate is very small ($\sim 2.13\,\mu\mathrm{m}$) and 
not compatible to standard waveguide technologies.
Thus, a new platform is required to achieve efficient QFC from the infrared to the UV spectral range in the quantum domain.
Potassium titanyl phosphate (KTP) is a material with a high photorefractive damage resistance. It allows not only the fabrication of short poling periods for QPM, but also the fabrication of high quality waveguides by rubidium ion exchange~\cite{Pysher2009}. An efficient QFC between a telecommunication and a red wavelength has been realized in rubidium doped periodically poled KTP (Rb:PPKTP) waveguide~\cite{Maring2014}.

Apart from the requirements concerning nonlinear coupling, the noise being added in a QFC device must not destroy the converted quantum state. Sources of detrimental noise have been attributed to Raman-scattering of the strong pump and non-phasematched spontaneous parametric downconversion (SPDC)~\cite{Pelc2011}. 
By using a pump wave that is spectrally well separated from in- and output, the effect of Raman-scattering in KTP~\cite{Kugel1988} can be neglected. Noise added by SPDC may however be present. But as shown in the case of a Rb:PPKTP waveguide converter~\cite{Maring2014} this noise does not prevent interfacing to a quantum memory, highlighting the potential of a single-stange Rb:PPKTP waveguide converter.

Among stationary qubit systems, a particularly interesting system for quantum information processing is a trapped ytterbium ion (Yb${}^{+}$) with its $S_\frac{1}{2} \rightarrow P_\frac{1}{2}$ dipole transition in the UV spectral region at $369.5\,\mathrm{nm}$~\cite{Olmschenk2010, Moehring2007}. It has been considered to be a good candidate to produce efficient light matter interaction~\cite{Maiwald2012} and entanglement of matter qubits~\cite{Trautmann2014}.
\begin{figure}[tb]
	\centering\includegraphics{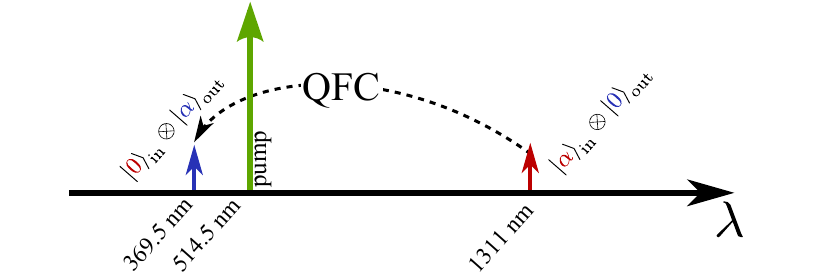} %[width=5.25in]
	\caption{Concept of direct quantum frequency upconversion by sum-frequency generation. A strong green pump is used to transduce a quantum state at a telecommunication-band input mode to a UV output mode.
					}
	\label{fig:concept}
\end{figure}%

In this paper we report on the tailored design and the classical characterization of a quantum interface
that allows for telecommunication-to-UV quantum frequency conversion, providing access to the Yb${}^{+}$ transition at $369.5\,\mathrm{nm}$. 
We present a detailed experimental analysis of the dispersion behavior and demonstrate first conversion efficiency measurements in a setup, which is compatible with quantum applications. By comparing our results to theoretical predictions we can evaluate the quality of feasible performance parameters. Thereby we provide
the conceptual and technical basis for a full quantum interface. 
\section{Concept}
Our interface is based on a second-order ($\chi^{(2)}$) SFG process in a rubidium doped PPKTP waveguide. 
A strong green pump at a wavelength of 514.5 nm is used in order to bridge the gap between the input state at $1311\,\mathrm{nm}$ and the target wavelength of the Yb${}^{+}$ dipole transition at $369.5\,\mathrm{nm}$ as shown in Fig.~\ref{fig:concept}. 

In order not to distribute the input state over several longitudinal modes, a CW single-mode pump laser is required, allowing to convert arbitrary input states within the process bandwidth. 
This pump fulfills the requirement of energy conservation, 
$\hbar\omega_\mathrm{in} + \hbar\omega_\mathrm{pump} = \hbar\omega_\mathrm{out}$,
where $\omega$ denotes the angular frequency of the optical fields at the respective wavelengths. 
At the same time its field amplitude $A_\mathrm{pump}$ scales the required coupling strength in the QFC-Hamiltonian~\cite{Kumar1990}
\begin{equation}
\hat{\mathcal{H}} = \mathrm{i}\hbar\kappa A_\mathrm{pump} \hat{a}^{\dagger}_\mathrm{out} \hat{a}_\mathrm{in} + \mathrm{h.c.},
\end{equation}%
where $\hat{a}$ is the photon annihilation operator of the in- and output mode and $\kappa$ is a coupling constant, dependent on the field overlap between the modes and their respective frequencies and propagation constants $\beta$.

Phasematching is attained by use of the quasi-phase\-matching (QPM) technique where a mismatch in the propagation constants $\Delta \beta$ is compensated by a periodic reversal of the ferroelectric domains, i.e. a domain grating of periodicity $\Lambda$, according to
\begin{equation} 
0 = \Delta\beta = \beta_\mathrm{out} - \beta_\mathrm{in} - \beta_\mathrm{pump} - \frac{2\pi}{\Lambda} .
	\label{eqn:pm}
\end{equation} 
In order to be able to use the highest second order nonlinear tensor element $d_{33}$, we employ a type-0-process, i.e. all fields are polarized along the optical axis. The short wavelength of $369.5\,\mathrm{nm}$ present in this process leads to a high mismatch of propagation constants and therefore a poling period below 3 $\mu\mathrm{m}$ is required in KTP. KTP allows, due to its anisotropic domain growth, such short periods. 
\section{Modeling}
\label{sec:model}
A prediction about the efficiency of the QFC in a Rb:PPKTP waveguide is obtained by modeling the effective refractive index of the waveguide modes and their respective field distributions. Although the material properties of KTP in the UV range are not comprehensively investigated in the literature, this approach allows us to provide a reliable model for the desired conversion process. 

The waveguide used in the presented work is produced by AdvR Inc. by means of potassium-rubidium ion exchange of z-cut, flux-grown KTP and subsequent periodic poling~\cite{Pysher2009}.  
It has a length of $9.6\,\mathrm{mm}$, the poling period is specified as $\Lambda = 2.535\,\mu\mathrm{m}$ and the waveguide has a nominal width of $2\,\mu\mathrm{m}$.
\begin{figure}[tb]
	\centering\includegraphics{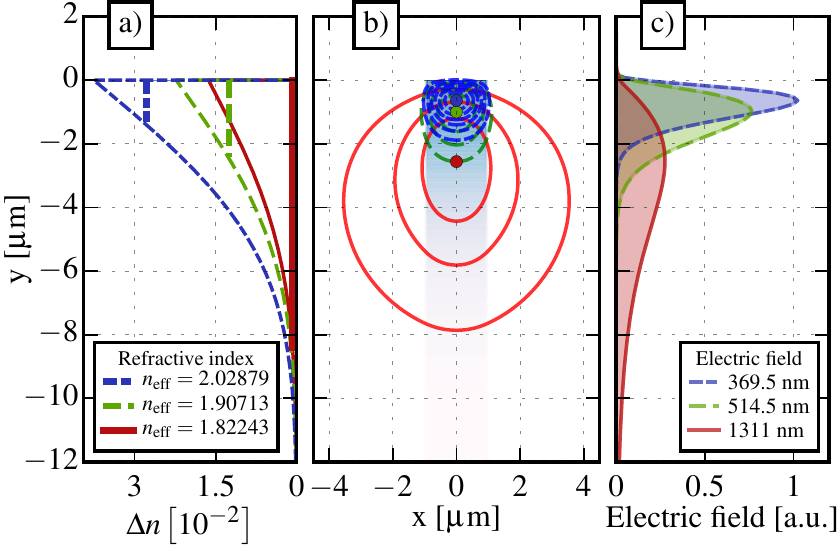} %[width=5.25in]
	\caption{Finite element simulation of modes in Rb:KTP waveguide including dispersive refractive index $n_\mathrm{bulk}$ and dispersive refractive index increase $\Delta n$.
	 a) Dispersive refractive index increase $\Delta n$ along the line $x=0\,\mu\mathrm{m}$. Vertical lines indicate the respective effective indizes for each of the modes.
		b) Lines of equal electric field strength of the fundamental modes in the rubidium doped waveguide region (shaded area). 
			c) Electric field of the respective fundamental modes along the line $x=0\,\mu\mathrm{m}$. 
	}
	\label{fig:model}
\end{figure}%
The ion exchange creates an $\mathrm{erfc}$-like Rb concentration profile~\cite{Bierlein1987} in depth and a box-like profile in lateral direction. In order to model the waveguide properties, we assume the Sellmeier equations of \textit{Kato and Takaoka}~\cite{Kato2002} for bulk KTP material and a refractive index profile according to the model presented by \textit{Callahan et. al.}~\cite{Callahan2014}. 
The refractive index profile in depth direction is depicted in Fig.~\ref{fig:model}a).

The propagation constants and electric field distributions at the respective wavelengths are calculated using a finite-element solver (\textit{RSoft FemSIM}). Assuming a penetration depth of $6\,\mu\mathrm{m}$~\cite{Callahan2014,Fedorova2015}, we calculate the respective effective indizes of the modes, as shown as vertical lines in Fig.~\ref{fig:model}a) and thus a theoretical poling period of $\Lambda_\mathrm{theo.} = 2.539\,\mu\mathrm{m}$ using equation~(\ref{eqn:pm}) and the relation between vacuum wavenumber $k$ and propagation constant, 
$\beta = k\cdot n_\mathrm{eff}$. 
This is in good agreement to the nominal poling period of the waveguide used in the remainder of this study.

The electric field distributions of the fundamental modes at the respective wavelengths are shown in Fig.~\ref{fig:model}b) and c).
According to this simulation, we expect our waveguide to support only a single mode per polarization at $\lambda_\mathrm{in} = 1311\,\mathrm{nm}$. Experimentally, however, we observe one higher mode in both TM and TE polarization. The expected number of guided modes in TM polarization is 9 and 29 at the wavelength of $\lambda_\mathrm{pump} = 514.5\,\mathrm{nm}$ and $\lambda_\mathrm{out} = 369.5\,\mathrm{nm}$, respectively. Because of the waveguide's asymmetry, a high number of mode combinations will have a non-vanishing overlap and can in principle contribute to the phasematching. The fundamental mode combination is favorable because of its rather high overlap to laser and fiber modes. The expected coupling efficiency of the infrared mode shown in Fig.~\ref{fig:model} to a Gaussian-shaped mode evaluates to $91.2\%$. 

The conversion efficiency for
a waveguide of length $L$ with negligible propagation loss in the regime of no pump depletion is~\cite{Roussev2006}
\begin{equation}
\eta  = {\sin ^2}\left( {\sqrt {{\eta _{\mathrm{nor}}}{P_\mathrm{pump}}} L} \right),
\label{eqn:conveff}	
\end{equation}
where $P_\mathrm{pump}$ is the pump power. It scales with the normalized efficiency 
\begin{equation}
{\eta _{\mathrm{nor}}} = \frac{8\pi^2}{c\cdot \varepsilon_0} \cdot \frac{d_{\mathrm{eff}}^2{\widetilde{\kappa} ^2}}{{{n_{\mathrm{eff}}^{\mathrm{(pump)}}}{n_{\mathrm{eff}}^{\mathrm{(in)}}}{n_{\mathrm{eff}}^{\mathrm{(out)}}}}}\cdot \frac{1}{\lambda_\mathrm{in} \lambda_\mathrm{out}} \ ,
\end{equation}
where $c$ is the speed of light and $\varepsilon_0$ the vacuum permittivity.
Here, we assume an effective non-linearity of $d_\mathrm{eff}= 16.65\cdot \frac{2}{\pi}\,\frac{\mathrm{pm}}{\mathrm{V}}$~\cite{Reshak2010} and use the calculated effective indices of refraction $n_{\mathrm{eff}}$.
The normalized overlap integral evaluates to 
$\widetilde{\kappa} = \int\int E_\mathrm{in}E_\mathrm{pump}E_\mathrm{out} \mathrm{d}x\mathrm{d}y = 135000\,\mathrm{m}^{-1}$,
where $E_\mathrm{in}$, $E_\mathrm{pump}$ and $E_\mathrm{out}$ are the three normalized mode fields of input, pump and output, shown in Fig.~\ref{fig:model}b) and c).
Thus, a pump power of $1460\,\mathrm{mW}$ is expected to be needed for complete conversion. 
Traces of the expected conversion efficiency are shown later in the paper in Fig.~\ref{fig:thedepl} where they are compared to measured data.
\section{Interface characterization}
\begin{figure}[tb]
	\centering\includegraphics{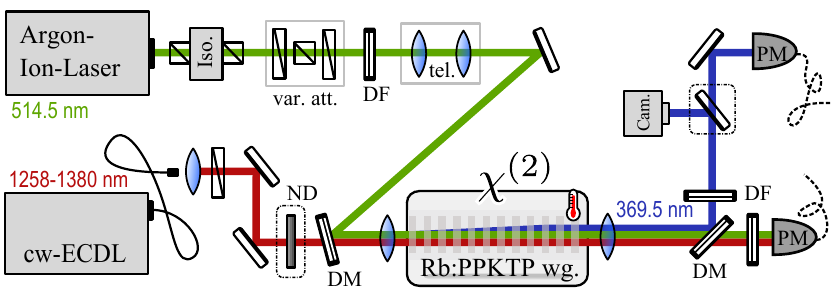} %[width=5.25in]
	\caption{Scheme of the experimental characterization setup. The following abbreviations are used; Iso.: Isolator, var.att.: variable attenuator, DF: dichroic filter, DM: dichroic mirror, tel.: telescope, ND: neutral density filter, PM: power meter.
					}
	\label{fig:setup}
\end{figure}

We characterize our interface using bright input light. The optical setup is depicted in Fig.~\ref{fig:setup}. 
The strong CW-pump power at $514.5\,\mathrm{nm}$ is provided by a single longitudinal and spacial mode Argon-Ion-Laser (\textit{Coherent Innova 90-C}). An optical isolator, consisting of two polarizing beam-splitters and a Faraday-rotator is used to prevent instabilities due to backreflections from the waveguide endface into the laser cavity. 
A combination of polarizing beam-splitter and half-wave-plates is used to control the optical power in the setup. 
Dichroic filters 
are needed to clean the beam from residual radiation, especially in the UV, produced by the laser's plasma discharge tube. 
A telescope is employed to manipulate the pump beam diameter and divergence for optimal coupling to the fundamental waveguide mode. 
The infrared input beam is produced by a fiber-coupled, tunable CW-external cavity diode laser (ECDL, \textit{Thorlabs Intun TL1300-B}). Its output is collimated by a $10\,\mathrm{mm}$-lens and overlapped with the green pump on a dielectric beam combiner. Both beams are coupled to the waveguide by a $20\times$ microscope objective.

The waveguide chip resides on a 6-axis adjustable stage and is thermoelectrically temperature controlled to $\pm 4\,\mathrm{mK}$ around room temperature. 
The generated UV light, together with the unconverted input light and the pump are coupled out of the waveguide with a $40\times$-microscope objective.

A dichroic mirror
reflects the UV light. Four successive bandpassfilters, centered at $370\,\mathrm{nm}$, 
with a cumulative optical density above 22 at the pump wavelength are used to filter out the remaining pump light. A Si photodiode based power meter 
with a sensitivity of $500\,\mathrm{pW}$ and a measurement uncertainty of $\pm5\,\%$ is used to monitor the converted light. Additionally, a CCD camera, 
adjustably placed in the beam, is used to record the nearfield intensity of the generated light.
\subsection{Modes and phasematching}
\begin{figure}[tb]
	\centering\hskip0.13pc\includegraphics{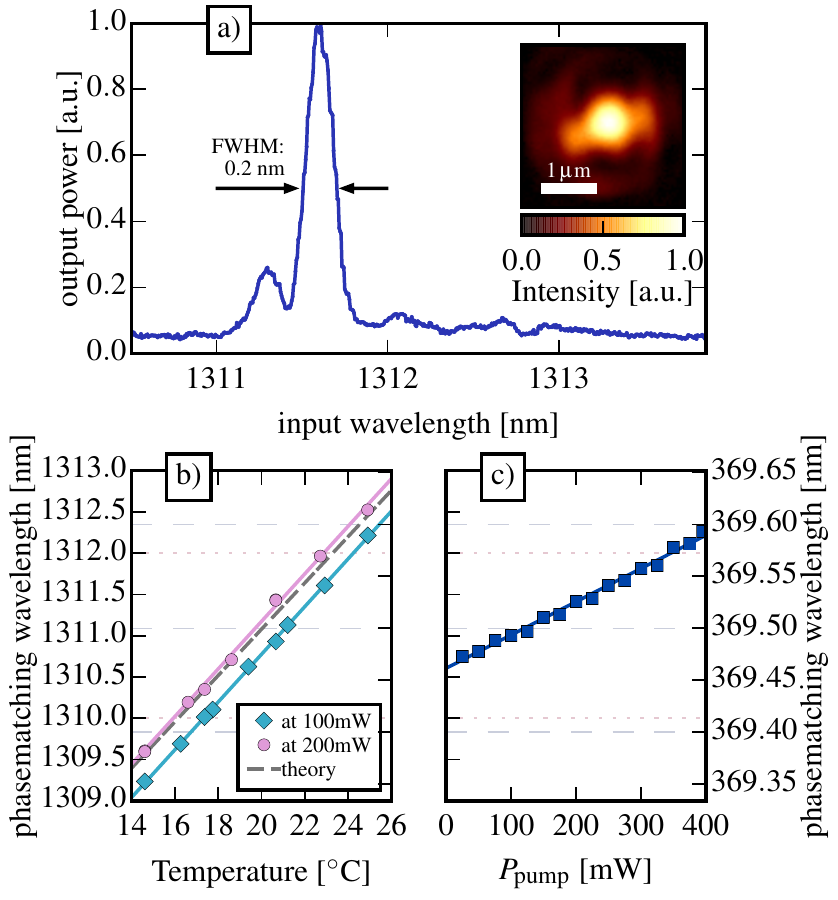} %
	\caption{Phasematching characteristics. 
		a) Phasematching curve, i.e. UV output power as function of input wavelength at stable temperature and pump power. Inset: Near field intensity profile of generated UV light in fundamental mode.
		b) Converter's temperature tunability, i.e. wavelength of perfect phasematching as function of externally controlled temperature.   
		c) Pump power induced phasematching shift. 
		Note that diagrams b) and c) share the same vertical scaling and thereby embody the energy conservation of sum frequency generation with a fixed pump at $514.5\,\mathrm{nm}$. 
		}
	\label{fig:pm}
\end{figure}

The phasematching condition of the fundamental mode combination is experimentally determined by iteratively adjusting the coupling of both, pump and input beam, to the respective fundamental waveguide mode and scanning the input wavelength whilst measuring the generated UV power and observing the spacial mode shape. 
Eventually, this procedure will lead to an optimized SFG-process in the fundamental mode combination. The SFG power as a function of the input wavelength at a constant pump power is depicted in Fig.~\ref{fig:pm}a). The full width at half maximum (FWHM) of this phasematching curve is $0.20(1)\,\mathrm{nm}$ while the theoretically predicted phasematching bandwidth is $0.185\,\mathrm{nm}$. This results in an effective interaction length of $8.9\,\mathrm{mm}$. The appearance of a pronounced sidelobe at shorter input wavelengths is consistent with the slightly reduced effective interaction length, resulting either from an inhomogeneous waveguide or poling period.
A minor flat contribution to the phasematching is observed over a much broader spectral range which we attribute to residual higher order mode's SFG-processes.

The near field intensity profile of the generated UV light is depicted in the inset of Fig.~\ref{fig:pm}a). 
The measured intensity FWHM is $0.94\,\mu\mathrm{m}$ in lateral and $0.77\,\mu\mathrm{m}$ in depth direction. This is in good agreement with the $1.10\,\mu\mathrm{m}$ and $0.82\,\mu\mathrm{m}$ predicted by our FEM simulation. 
 
The phasematching wavelength of the frequency conversion process is tunable with temperature. In order to model this, we assume perfect phasematching at $T = 20{\,}^{\circ}\mathrm{C}$ and $\lambda_\mathrm{in} = 1311\,\mathrm{nm}$ and calculate the phasematching condition according to the temperature dependent index for bulk material~\cite{Kato2002}. The expected change of the phasematching wavelength $\Delta\lambda_\mathrm{in}$ at the input as a function of the temperature difference $\Delta T$ is $\Delta\lambda_\mathrm{in} / \Delta T = 0.28\,\mathrm{nm}/\mathrm{K}$. 
This behaviour is plotted as a dashed line in Fig.~\ref{fig:pm}b). Experimentally we measure $\Delta\lambda_\mathrm{in} / \Delta T = 0.29(1) \mathrm{nm}/\mathrm{K}$, as depicted in the same diagram for two different pump powers.

In general, we observe a strong dependence of the phasematching wavelength on the pump power, which is shown in Fig.~\ref{fig:pm}c). The shift in phasematching wavelength $\Delta\lambda_\mathrm{in}$ behaves linearly with the external pump power $P_\mathrm{pump}$ according to 
$\Delta\lambda_\mathrm{in}/\Delta P_\mathrm{pump} = 4.03(9)\,\mathrm{pm}/\mathrm{mW}$. This shift is the reason for the rather complicated identification of the phasematching described above. 
However, once a phasematching condition for a given pump power and temperature is determined experimentally, the converter's full tuning behavior is predictable by the two given values of $\Delta\lambda_\mathrm{in} / \Delta T$ and $\Delta\lambda_\mathrm{in}/\Delta P_\mathrm{pump}$.

Combining those values, the pump power dependent phasematching shift may be explained by a local heating of the waveguide according to $\Delta T/\Delta P_\mathrm{pump} = 13.8(6)\,\mathrm{K}/\mathrm{W}$ due to absorption of the green pump light.
\subsection{Conversion efficiency}
We define the conversion efficiency $\eta$ as the number of converted photons exiting the nonlinear crystal $\left\langle N\right\rangle_\mathrm{out}$ divided by the number of input photons in front of it $\left\langle N\right\rangle_\mathrm{in}$,
\begin{equation}
	\eta = \frac{\left\langle N\right\rangle_\mathrm{out} }{\left\langle N\right\rangle_\mathrm{in}} = 
						\frac{\lambda_\mathrm{out}}{\lambda_\mathrm{in}} \cdot\frac{P_\mathrm{out}}{P_\mathrm{in}} .
\end{equation}
In the case of no losses, this should equal the depletion efficiency 
\begin{equation}
\eta_\mathrm{depl.}  = 1- \frac{\left\langle N\right\rangle_\mathrm{depl.} }{\left\langle N\right\rangle_\mathrm{in}} = 1 - \frac{P_\mathrm{in, depl.}}{P_\mathrm{in}}
	\label{eqn:depletion}
\end{equation}
where $P_\mathrm{in, depl.}$ is the optical power of the input field at the end of the nonlinear interaction length.

The measurement of the conversion efficiency is performed as follows. 
We keep the waveguide chip at a constant temperature of $T=20.6^{\circ}\mathrm{C}$, attenuate the input light to $P_\mathrm{in} \cong 20\,\mu\mathrm{W}$ and vary the pump power between $0$ and $400\,\mathrm{mW}$. Higher powers of $\geq 550\,\mathrm{mW}$ have shown to cause significant damage to the crystal.
For each pump power in use, we adapt the input wavelength to the phasematching according to the pump power dependent relation described above.
The system is then given enough time to equilibrate in order to avoid hysteretic behavior.
The optical power around $369.5\,\mathrm{nm}$ is recorded using a power meter. The optical power remains stable on a scale between the power meter's bandwidth ($\sim 10\,\mathrm{Hz}$) and half an hour. 
Each data point is recorded as the average power over 5 seconds and the fluctuations of the optical signal are much smaller than the power meter's measurement uncertainty.
At a pump power of $P_\mathrm{pump} = 200\,\mathrm{mW}$ and an input power of $P_\mathrm{in} = 22.1\,\mu\mathrm{W}$ in front of the incoupling objective, we measure 
a UV power of $P_\mathrm{out} = 980\,\mathrm{nW}$. Accounting for the transmission of the incoupling objective ($69.6\,\%$),  
the outcoupling ($51.7\,\%$),
and the transmission of the filters and optical components up to the detection ($62.7\,\%$),  
an external conversion efficiency of $\eta_\mathrm{ext.} = 5.5\,\%$ is obtained. We call this measure 'external', because we explicitly neither account for the coupling efficiency to the fundamental mode, nor the waveguide absorption and scattering losses, as both cannot be quantified accurately.
\begin{figure}[tb]
	\centering\includegraphics{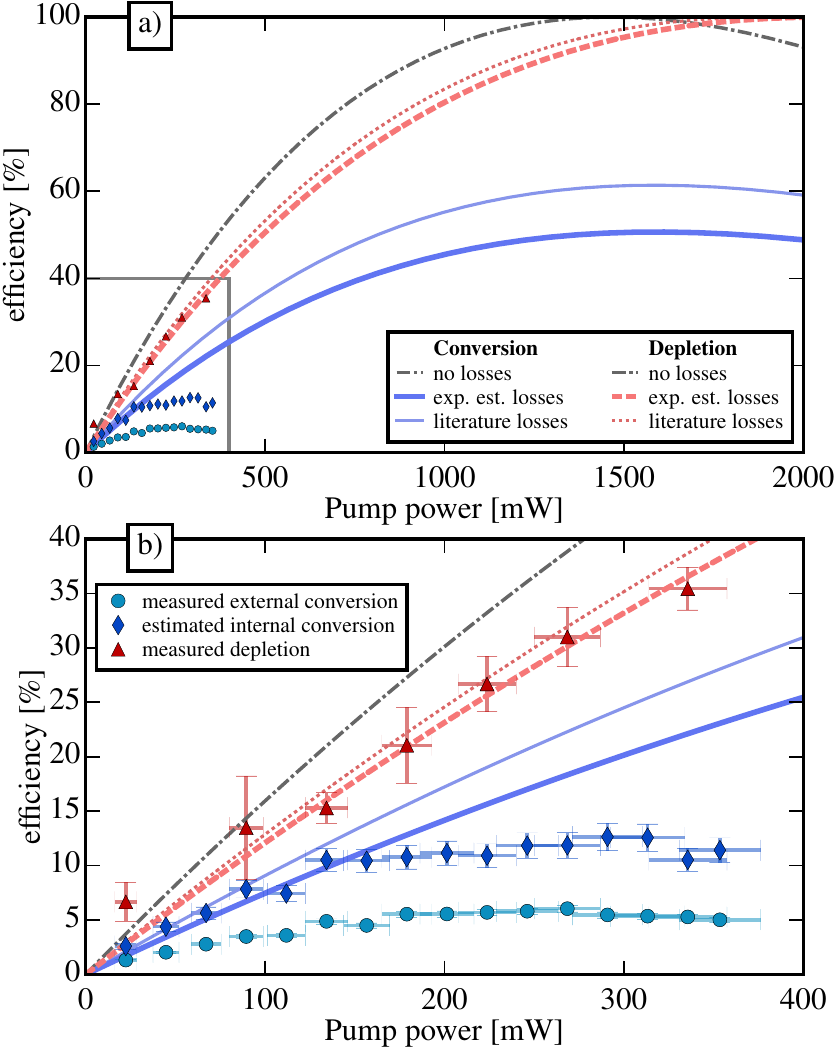} %
	\caption{Efficiency of the sum-frequency process, experimental values compared to theoretical predictions. b) is a zoom into a) for clarification.
	A lossless process (dash-dotted line) would show identical conversion and depletion efficiency. In the case of losses, depletion (dashed lines) and conversion (solid lines) efficiency differ.
	Measured (triangles) and theoretically predicted (dashed lines) depletion efficiency are in good agreement. 
	The measured external conversion efficiency (circles) reaches $5.5\,\%$. The estimated internal conversion efficiency (diamonds) also shows a good agreement to theoretical predictions 
	for internal pump powers up to $150\,\mathrm{mW}$. For higher pump powers, clear deviation from the model is observed.
	}
	\label{fig:thedepl}
\end{figure}%
The external conversion efficiency is plotted as circles in Fig.~\ref{fig:thedepl}, where we also account for the imperfect transmission of the incoupling objective ($89.5\,\%$) at the pump wavelength. 

In the ideal case of perfect mode matching and negligible losses, the conversion efficiency, as well as the depletion, should follow the dash-dotted curve in Fig.~\ref{fig:thedepl} which is the theoretically predicted conversion efficiency for the Rb:PPKTP-waveguide considered in section~\ref{sec:model} using equation~(\ref{eqn:conveff}).

The more realistic case is calculated by solving the nonlinear coupled amplitude equations, including losses, for perfect phasematching (see e.g.~\cite{Roussev2006}). This case is shown as the thin curves in Fig.~\ref{fig:thedepl} (\textit{literature losses}).
We include losses of $\alpha_\mathrm{pump} = 0.7\,\,\mathrm{dB}/\mathrm{cm}$ at the pump,
$\alpha_\mathrm{in} = 0.2\,\mathrm{dB}/\mathrm{cm}$ at the in-, 
and ${\alpha_\mathrm{out} = 4.34\,\mathrm{dB}/\mathrm{cm}}$ at the output wavelength.  
Here, we estimate the pump loss from the values reported in reference~\cite{Pysher2009} and the input loss from the lowest loss we measured in Rb:KTP waveguides at that wavelength. The exact values of both are negligible compared to the
high loss at the output wavelength assumed here. The latter relies on the fact that $370\,\mathrm{nm}$ already lies close to the band edge for z-polarized light in KTP~\cite{Hansson2000}. 

Because the loss for y-polarized light is much lower at that wavelength, the relative transmission between TM- and TE- polarized laser light can provide an estimate for the expected waveguide losses at $370\,\mathrm{nm}$. 
This relative transmission is measured to be $33.6\,\%$ ($\alpha_\mathrm{out} = 6.3\,\mathrm{dB}/\mathrm{cm}$). We call this model \textit{experimentally estimated} and depict it using thick lines in Fig.~\ref{fig:thedepl}.

In all cases we would expect a monotonically increasing conversion efficiency up to at least $25\,\%$ within the available pump power range. In contrast, the measured conversion efficiency saturates and decreases for pump powers above $300\,\mathrm{mW}$.
For a better comparison to the model, we estimate the internal conversion efficiency, i.e. the conversion efficiency between in- and output light inside the respective waveguide mode. This is shown as diamond shaped dots in Fig.~\ref{fig:thedepl}. At a pump power of $P_\mathrm{pump} = 200\,\mathrm{mW}$ in front of the incoupling, we estimate an internal conversion efficiency of $\eta_\mathrm{int} = 10.5\,\%$ by assuming fresnel losses at the uncoated input facet and modematching of $50.7\,\%$.

Naturally, the expected depletion and conversion efficiency differ substantially, when assuming a lossy system. The depletion would reach up to $40\,\%$ in the  
considered case. This is verified experimentally by adjusting the system to optimal conversion and chopping the pump light with frequencies of 20, 100, and 500 Hz. The modulation of the transmitted input light is recorded with a fast photodiode. This provides a direct measure of the depletion efficiency according to equation~(\ref{eqn:depletion}). The depletion efficiency is plotted as triangles in Fig.~\ref{fig:thedepl}. 
It does not depend on the chopping frequency. Fluctuation of the data points due to amplified noise in the photocurrent and data evaluation uncertainties are accounted for in the shown errorbars. Within those errorbars, the measured and theoretically predicted depletion values are in good agreement. 

Even in the pump power range in which we observe a reduced conversion efficiency the depletion follows the expected behavior. A possible 
mechanism leading to a divergence of depletion and conversion is the so called green induced infrared absorption (GRIIRA)~\cite{Wang2004}, frequently observed in KTP. We quantify this effect to account for at most $6.6\,\%$ depletion by tuning the input wavelength away from the phasematching condition and repeating the above measurement.

Our converter therefore behaves according to the model given by the nonlinear coupled amplitude equations up to a maximum external pump power of $200\,\mathrm{mW}$. At higher pump powers a clear deviation from this model is observed which needs to be subject to further investigation. %
\section{Conclusion and Outlook}
In conclusion, we have implemented a telecommunication-to-UV frequency converter, compatible with the requirements of QFC.
It is based on a second-order nonlinear sum-frequency process in a periodically poled Rb:PPKTP waveguide.
Using a fixed single-mode pump at $514.5\,\mathrm{nm}$, the device is tunable by temperature and/or pump power and thereby allows to interface the telecommunications O-band and the Yb${}^{+}$ transition at $369.5\,\mathrm{nm}$ with an external (internal) efficiency of $\eta_\mathrm{ext} = 5.5\,\%$ ($\eta_\mathrm{int} = 10.5\,\%$).
The internal efficiency is limited by the material absorption at the output wavelength, the mode overlap between the very distinct wavelengths of pump, in- and output field and the short interaction length of $\leq 1\,\mathrm{cm}$, combined with a saturation of the conversion efficiency above $200\,\mathrm{mW}$ pump power. For lower pump powers the nonlinear coupled amplitude equations of the three wave mixing process are shown to provide a good model for the device efficiency. 
The external efficiency is limited by coupling losses. Proper antireflection coatings and a specifically designed achromatic coupling lens would significantly improve the coupling into the desired waveguide mode.
 
In future implementations, the use of a pulsed pump might be beneficial for two reasons. First, it reduces average power while maintaining peak power needed for conversion. This would reduce the shift of phasematching observed in this paper and possibly reduce optical damage.
Second, a pulsed pump would allow for temporal filtering in a later stage of the experiment and thereby has the potential to reduce possible noise counts at the converter's output.
 
In order for our converter to be an eligible QFC device, it needs to retain its conversion efficiency on the single photon level and its noise characteristics need to be carefully investigated. 
While the strong green pump in our converter is placed spectrally far away from in- and output, reducing the effect of added noise by Raman-scattering to a minimum, its short wavelength opens the possibility for generation of SPDC photons into the input mode.
After having provided the classical characteristics of our telecommunication-to-UV frequency converter in this paper, our next step thus needs to be its quantum noise characterization.

Despite the moderate conversion efficiency, we expect our device to be highly useful for quantum information tasks involving direct access to trapped ion systems. Especially, it allows to interface and thus integrate trapped ions at UV wavelengths into optical fiber network architectures. 
Note e.g. that compared to a direct transmission of UV light in a fiber where losses around $0.1\,\mathrm{dB}/\mathrm{m}$ appear as a reasonable number~\cite{Colombe2014}, the use of a 5\% ($-13\,\mathrm{dB}$) efficient converter may be beneficial already for distances above $130\,\mathrm{m}$.
Moreover, our conversion opens the possibility to generate non-classical states of light at UV wavelengths by using cascaded spontaneous parametric downconversion together with SFG. 
Those quantum states are needed for studying the coupling between genuine quantum light and isolated single particle excitations with UV transitions energies.
\acknowledgement{We thank Harald Herrmann for helpful discussions and also the reviewers for useful comments, contributing to improve the manuscript. We acknowledge financial support provided by the German Bundesministerium f\"ur Bildung und Forschung within the \textit{QuOReP} and \textit{Q.com-Q} framework.}
%
%
% BibTeX users use
\bibliography{refs}
\bibliographystyle{osajnl} % % spphys not working
\end{document}